\documentclass[12pt]{article}
 \usepackage{epsfig}
 \def\be{\begin{equation}}
 \def\ee{\end{equation}}
 \def\bea{\begin{eqnarray}}
 \def\eea{\end{eqnarray}}
 \usepackage{graphicx}

 \catcode`\@=11
 \def\lsim{\mathrel{\mathpalette\@versim<}}
 \def\gsim{\mathrel{\mathpalette\@versim>}}
 \def\@versim#1#2{\vcenter{\offinterlineskip
 \ialign{$\m@th#1\hfil##\hfil$\crcr#2\crcr\sim\crcr } }}
 \catcode`\@=12

 \catcode`@=12
\begin{document}
 \thispagestyle{empty}
 \begin{flushright}
 UCRHEP-T626\\
 November 10, 2023\
 \end{flushright}
 \vspace{0.6in}
 \begin{center}
 {\LARGE \bf Softly Broken Hidden Symmetry in\\ 
Every Renormalizable Field Theory\\}
 \vspace{1.5in}
 {\bf Ernest Ma\\}
 \vspace{0.1in}
{\sl Department of Physics and Astronomy,\\ 
University of California, Riverside, California 92521, USA\\}
 \vspace{1.2in}

{\it (in memory of Eileen)\\}
\end{center}

\begin{abstract}\
It is pointed out that every renomalizable field theory has a symmetry 
which is hidden in plain sight.  In all practical cases, it is also broken 
softly, either explicitly or spontaneously.  The soft explicit breaking mass 
terms may be assumed naturally small compared to the scalar mass-squared 
terms. Implications for extensions of the standard model are discussed.
New left-right model of two-loop radiative Dirac neutrino mass is proposed.
\end{abstract}

\newpage
\baselineskip 24pt

\noindent \underline{\it Introduction}~:~ 
Renormalizable quantum field theory has long been the tool to study particle 
physics.  It uses a Lorentz invariant Lagrangian to describe the propagations 
and interactions of fundamental scalar, fermion, and vector gauge particles 
in precise well-defined calculations.  The well-konwn standard model (SM) of 
quarks and leptons is represented by such a Lagrangian, and studies of what 
physics lies beyond are mostly based on this approach as well.

\noindent \underline{\it Observation}~:~
Consider a scalar field $\phi$, a left-handed fermion field $\psi_L$, 
a right-handed fermion field $\psi_R$, and a vector gauge field $A_\mu$. 
Under the discrete $Z_4$ symmetry, let
\begin{equation}
A_\mu \sim 1, ~~~ \phi \sim -1, ~~~ \psi_L \sim i, ~~~ \psi_R \sim -i,
\end{equation}
then it is obvious that all allowed terms in any renormalizaable Lagrangian 
are invariant under it, except the dimension-three terms
\begin{equation}
\phi^3, ~~~ \bar{\psi}_L \psi_R, ~~~ \psi_L \psi_L, ~~~ \psi_R \psi_R.
\end{equation}
In the presence of one or more of these terms, $Z_4$ is explicitly broken 
to $Z_2$, under which
\begin{equation}
A_\mu,\phi \sim 1, ~~~ \psi_{L,R} \sim -1,
\end{equation}
which is just the well-known statement that it takes two fermions to 
make a boson.  Nothing of substance seems to have been gained.

\noindent \underline{\it Insight}~:~ 
The terms of Eq.~(2) all have mass dimensional couplings.  Their counterterms 
from all the interactions of the theory are linearly proportional to these 
couplings themselves.  Hence their absence guarantees the preservation of 
the $Z_4$ symmetry.  It may thus be argued that such terms should be small. 

Consider now the possible spontaneous breaking of $Z_4$, i.e. 
$\langle \phi \rangle \neq 0$.  This is normally triggered by the 
\underline{allowed} quadratic and quartic scalar terms.  Hence the resulting 
dimension-three terms are normally of the magnitude of the original 
quadratic scalar mass term.  However, this does not invalidate the statement 
that the \underline{explicit} $Z_4$ breaking terms may be chosen to be small. 

As an explicit example, consider a single real scalar $\phi$.  Its potential 
is
\begin{equation}
V = {1 \over 2} m^2 \phi^2 + {1 \over 6} \mu \phi^3 + {1 \over 24} \lambda 
\phi^4.
\end{equation}
It is clear that the $\mu$ term may be chosen to be small compared to 
$|m|$.  If $m^2 < 0$, then $\langle \phi \rangle = v$ is given by
\begin{equation}
m^2 + {1 \over 2} \mu v + {1 \over 6} \lambda v^2 = 0.
\end{equation}
Let $\phi = \phi' + v$, then
\begin{equation}
V = {1 \over 2} m^2 (\phi'+v)^2 + {1 \over 6} \mu (\phi'+v)^3 + {1 \over 24} 
\lambda (\phi'+v)^4.
\end{equation}
Again, $|\mu| << |m|$ is a natural assumption because shifting a scalar field 
by a constant does not change the structure of the Lagrangian.  However, the 
$(\phi')^3$ coupling is now $(\lambda v + \mu)/6$ which is of order $|m|$. 
This means that whereas $|\mu| << v$ is technically correct, it has no 
practical importance in this simple example.  However, it does illustrate 
how explicit small breaking is maintained in the presence of large 
spontaneous breaking.  It will also become relevant where the specific 
spontaneous breaking is triggered by an explicit breaking term as in other 
examples to be shown below.

\noindent \underline{\it Stnadard Model}~:~ 
The SM is based on the gauge symmetry $SU(3) \times SU(2) \times U(1)$. 
As such, its Lagrangian has no dimension-three terms before symmetry breaking. 
Once its Higgs doublet $\Phi = (\phi^+,\phi^0)$ acquires a vacuum expectation 
value, fermion mass terms such as that of the top quark and a trilinear Higgs 
coupling do appear and are of order the Higgs mass.  This is expected, but 
since there are no \underline{explicit} $Z_4$ breaking dimension-three terms, 
there is no small mass parameter to be discussed in this case. 

Because of the chosen particle content of the standard model under its gauge 
symmetry, two additional (global) symmetries appear, i.e. baryon number $B$ 
and lepton number $L$.  The universal symmetry $Z_4$ is then given in 
this case by
\begin{equation}
Z_4 = (i)^{(B-L+4Y)}
\end{equation}
as shown in Table 1.
\begin{table}[tbh]
\centering
\begin{tabular}{|c|c|c|c|c|c|c|}
\hline
particle & $SU(3)_C$ & $SU(2)_L$ & $Y$ & $B$ & $L$ & $Z_4$ \\
\hline
gluon & 8 & 1 & 0 & 0 & 0 & 1 \\
$W^{\pm},Z,A$ & 1 & 3,1 & 0 & 0 & 0 & 1 \\
\hline
$(u,d)_L$ & 3 & 2 & 1/6 & 1/3 & 0 & $i$ \\ 
$u_R,d_R$ & 3 & 1 & 2/3,$-1/3$ & 1/3 & 0 & $-i$ \\ 
$(\nu,e)_L$ & 1 & 2 & $-1/2$ & 0 & 1 & $i$ \\ 
$e_R$ & 1 & 1 & $-1$ & 0 & 1 & $-i$ \\
\hline
$(\phi^+,\phi^0)$ & 1 & 2 & 1/2 & 0 & 0 & $-1$ \\
\hline
\end{tabular}
\caption{$Z_4$ realization of the standard model.} 
\end{table}

\noindent \underline{\it Neutrino Mass with Higgs Triplet}~:~ 
Adding the Higgs triplet $\xi = (\xi^{++},\xi^+,\xi^0)$ to the SM generates 
two extra terms: the dimension-4 term $f_{ij} \xi^0 \nu_i \nu_j + ...$ which 
obeys $Z_4$ and the dimension-3 term $\mu \bar{\xi}^0 \phi^0 \phi^0 + ...$ 
which violates it explicitly.  However, their coexistence now also breaks 
lepton number $L$, but there are two choices as shown in Table 2.
\begin{table}[tbh]
\centering
\begin{tabular}{|c|c|c|c|}
\hline
interaction & $L=-2$ & $L=0$ & $Z_4$ \\
\hline
$f_{ij} \xi^0 \nu_i \nu_j$ & yes & no & yes \\
$\mu \bar{\xi}^0 \phi^0 \phi^0$ & no & yes & no \\
\hline
\end{tabular}
\caption{$L$ assigments of scalar triplet.} 
\end{table}

The conventional choice is $L=-2$ in agreement with the universal symmetry 
$Z_4$. Let $\langle \phi^0 \rangle = v_0$ and assuming that $m^2_\xi >> v_0^2$, 
it is well-known that~\cite{ms98} 
\begin{equation}
v_1 = \langle \xi^0 \rangle \simeq {-\mu v_0^2 \over m^2_\xi}.
\end{equation}
This is important because $v_1$ is now proportional to the explicit $Z_4$ 
breaking $\mu$ term and is much smaller than $v_0$.  The new trilinear 
coupling of the $\bar{\xi}^0 hh$ term is $(\mu + \lambda_3 v_1)/2$ and 
remains small in contrast to the previous simple example.  In the above, 
$h=\sqrt{2}Re(\phi^0)$ is the SM Higgs boson and $\lambda_3$ comes from the 
$\lambda_3 (\bar{\xi}^0 \xi^0)(\bar{\phi}^0\phi^0)$ term.  Furthermore, 
neutrino masses are proportional to $v_1$ and are thus desirably small.

If $L=0$ is assumed for $\xi$, then the dimension-4 term 
$f_{ij} \xi^0 \nu_i \nu_j$ breaks $L$, whereas the dimension-3 term 
$\mu \bar{\xi}^0 \phi^0 \phi^0$ is allowed.  In this case, $v_1$ is not 
expected to be small by itself, whereas $f_{ij}$ should be very small for 
obtaining realistic neutrino masses. In pratice, $v_1$ contributes to the 
parameter
\begin{equation}
\rho = {M_W^2 \over M_Z^2 \cos^2 \theta_W} = 1.0002 \pm 0.0009.
\end{equation}
It adds $2(g^2/\cos^2 \theta_W)v_1^2$ to $M_Z^2$ and $g^2v_1^2$ to $M_W^2$, 
hence $\rho < 1$ is predicted.  Allowing two standard deviations of the 
data, this means that $v_1^2/v_0^2 < 0.0008$.  This argues again for the case 
$L=-2$ where $v_1$ is naturally small.

\noindent \underline{\it Neutrino Mass with Fermion Singlet}~:~ 
Adding the fermion singlet $N_R$ to the SM generates two extra terms: the 
dimension-4 term $f_N \bar{N}_R \nu_L \phi^0$ which obeys $Z_4$ and the 
dimension-3 term $(m_N/2) N_R N_R$ which violates it.  Again lepton number 
is broken with two choices as shown in Table 3. 
\begin{table}[tbh]
\centering
\begin{tabular}{|c|c|c|c|}
\hline
interaction & $L=1$ & $L=0$ & $Z_4$ \\
\hline
$f_N \bar{N}_R \nu_L \phi^0$ & yes & no & yes \\
$(m_N/2) N_R N_R$ & no & yes & no \\
\hline
\end{tabular}
\caption{$L$ assigments of fermion singlet.} 
\end{table}

What is the conventional choice in this case? The common name for $N_R$ is 
in fact the right-handed neutrino, implying that it is a lepton and should 
have $L=1$.  If so, why is $m_N$ always assumed to be large, as required by 
the seesaw mechanism? The usual answer is that $N_R$ is a gauge singlet 
and its mass is arbitrary.  However, with the realization that $m_N$ violates 
explicitly the universal symmetry $Z_4$, it should be small compared to 
the Higgs mass for example, and the seesaw mechansim does not work so well.

The other choice $L=0$ means that $m_N$ may indeed be large and the 
conventional seesaw mechanism prevails.  Note however that the basis of 
the universal $Z_4$ symmetry is that it applies to all dimension-4 
terms.  Therefore the price to pay for the usual seesaw mechanism is to 
assume a symmetry, i.e. $L=0$ for $N_R$, which is violated by a dimension-4 
term, i.e. $f_N \bar{N}_R \nu_L \phi^0$.

If $B-L$ gauge symmetry is imposed with $N_R \sim -1$, then the addition of 
a Higgs scalar $\chi$ with $B-L=2$ generates a mass for $N_R$ from 
$\langle \chi \rangle \neq 0$.  This is again unsuppressed because it comes 
from $Z_4$ allowed quadratic and quartic scalar terms.  In fact, there is no 
\underline{explicit} $Z_4$ breaking term in this case, and 
$m_D << m_N$ in the seesaw formula $m_\nu = m_D^2/m_N$ is 
maintained by the breaking of $B-L$ to the SM.  This version  
is thus consistent with the $Z_4 \to Z_2$ argument.

\noindent \underline{\it Neutrino Mass from Dark Matter}~:~ 
To accommodate dark matter and to link it with neutrino mass, the scotogenic 
model~\cite{m06} was invented.  The three fermion singlets $N_R$ as well as 
a second scalar doublet $\eta = (\eta^+,\eta^0)$ are odd under a dark 
$Z_2$ symmetry which is unbroken.  The SM neutrinos then obtain one-loop 
radiative masses through the Yukawa term 
$\bar{N}_R (\nu_L \eta^0 - l_L \eta^+)$ and the Majorana mass $m_N$ of 
$N_R$.  The original application of this idea assumes $m_\eta << m_N$. 
Recognizing the universal $Z_4$ symmetry, the more acceptable choice 
should be $m_N << m_\eta$, as studied in Ref.~\cite{m12}.  This implies 
light dark matter has masses and mixing pattern directly related to 
those of neutrinos.

\noindent \underline{\it New Left-Right Model of Radiative Dirac Neutrino 
Mass}~:~ 
As a new application of the universal $Z_4$ symmetry, consider a left-right 
model under $SU(3)_C \times SU(2)_L \times SU(2)_R \times U(1)_Y$ as shown in 
Table 4.
\begin{table}[tbh]
\centering
\begin{tabular}{|c|c|c|c|c|}
\hline
particle & $SU(3)_C$ & $SU(2)_L$ & $SU(2)_R$ & $U(1)_Y$ \\
\hline
$(u,d)_L$ & 3 & 2 & 1 & 1/6 \\ 
$u_R,d_R$ & 3 & 1 & 1 & 2/3,$-1/3$ \\ 
\hline
$(U,D)_R$ & 3 & 1 & 2 & 1/6 \\ 
$U_L,D_L$ & 3 & 1 & 1 & 2/3,$-1/3$ \\ 
\hline
$(\nu,e)_L$ & 1 & 2 & 1 & $-1/2$ \\ 
$e_R$ & 1 & 1 & 1 & $-1$ \\
\hline
$(\nu,E)_R$ & 1 & 1 & 2 & $-1/2$ \\ 
$E_L$ & 1 & 1 & 1 & $-1$ \\
\hline
$\Phi_L=(\phi_L^+,\phi_L^0)$ & 1 & 2 & 1 & 1/2 \\
$\Phi_R=(\phi_R^+,\phi_R^0)$ & 1 & 1 & 2 & 1/2 \\
\hline
\end{tabular}
\caption{Particle content of new left-right model.} 
\end{table}
The mass terms $\bar{u}_R U_L, \bar{d}_R D_L, \bar{e}_R E_L$ are allowed under 
the gauge symmetry, but are explicit soft terms which break $Z_4$. They will 
be assumed small.  Hence the SM quarks and leptons mix only very slightly 
with their heavy mirror counterparts.  The gauge symmetries $SU(2)_{L,R}$ are 
broken respectively by $\Phi_{L,R}$, resulting in two physical Higgs scalars 
$h_{L,R} = \sqrt{2} Re(\phi^0_{L,R})$.  The generic $2 \times 2$ mass matrix 
linking $(\bar{q}_L,\bar{Q}_L)$ to $(q_R,Q_R)$ is of the form
\begin{equation}
{\cal M}_{qQ} = \pmatrix{m_q & 0 \cr \tilde{m} & m_Q},
\end{equation}
where $m_q$ comes from $\langle \phi^0_L \rangle$, $m_Q$ from $\langle \phi^0_R 
\rangle$ and $\tilde{m}$ from the soft $Z_4$ breaking term $\bar{Q}_L q_R$. 
The zero occurs because there is no scalar bidoublet. 
It is diagonalized on the left by $\theta_L \simeq \tilde{m} m_q/m_Q^2$ and 
on the right by $\theta_R \simeq \tilde{m}/m_Q$.  The $W_L-W_R$ mixing is 
induced in one loop by this fermion connection as shown for example in Fig. 1.
\begin{figure}[htb]
\vspace*{-5cm}
\hspace*{-3cm}
\includegraphics[scale=1.0]{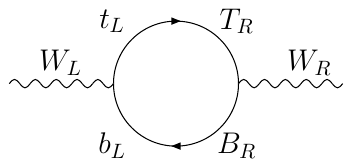}
\vspace*{-19.5cm}
\caption{$W_L-W_R$ mixing.}
\end{figure}

This contribution is finite and proportional to $\tilde{m}_{tT} \tilde{m}_{bB}$. 
Let the full contribution be denoted by $\tilde{m}^2_{eff}$, then $\nu_L$ is 
connected to $\nu_R$ in two loops as shown in Fig. 2.
\begin{figure}[htb]
\vspace*{-5cm}
\hspace*{-3cm}
\includegraphics[scale=1.0]{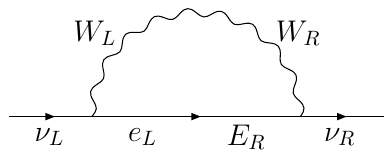}
\vspace*{-21.5cm}
\caption{Radiative Dirac neutrino mass.}
\end{figure}

A rough estimate yields
\begin{equation}
{m_\nu \over m_e} \simeq {g_L g_R \over 16 \pi^2} {\tilde{m}_{eE} \over m_E} 
{\tilde{m}^2_{eff} \over m^2_{W_R}},
\end{equation}
which may be of order $10^{-2} \times 10^{-2} \times 10^{-4} = 10^{-8}$ or less. 
With all three charged-lepton masses, realistic neutrino masses may be 
obtained.

If $\tilde{m}$ in Eq.~(10) is chosen to be large instead of small, then 
$u_R,d_R,e_R$ should be swapped with $U_R,D_R,E_R$ as  
in models of seesaw masses for all quarks and leptons.  
In the absence of a neutral fermion singlet (which is also the case here), 
this becomes the model of Ref.~\cite{bhst22}.  For an updated discussion 
of the many phenomenological constraints in this conventional choice, see 
Ref.~\cite{d24}.  As it is, the model discussed here is safe because the 
mixing between the SM and its mirror counterpart is always suppressed. 
For one-loop left-right realizations of Dirac neutrino mass, see 
Refs.~\cite{m88,m89,ms18}.

\noindent \underline{\it Conclusion}~:~
In model building beyond the SM, the universal $Z_4 \to Z_2$ 
\underline{explicit} soft breaking 
of any renormalizable Lagrangian could be a guide to distinguish the 
hierarchical choices of parameters that one encounters.  Which path it leads 
could of course only be proven by future experiments.

\noindent \underline{\it Acknowledgement}~:~
This work was supported in part by the U.~S.~Department of Energy Grant 
No. DE-SC0008541.  

\baselineskip 18pt
\bibliographystyle{unsrt}

\begin{thebibliography}{99}
\bibitem{ms98} E. Ma and U. Sarkar, Phys. Rev. Lett. {\bf 80}, 5716 (1998).
\bibitem{m06} E. Ma, Phys. Rev. {\bf D73}, 077301 (2006).
\bibitem{m12} E. Ma, Phys. Lett. {\bf B717}, 235 (2012).
\bibitem{bhst22} K. S. Babu, X.-G. He, M. Su, and A. Thapa, JHEP {\bf 2208}, 
140 (2022).
\bibitem{d24} R. Dcruz, Nucl. Phys. {\bf B1001}, 116519 (2024). 
\bibitem{m88} R. N. Mohapatra, Phys. Lett. {\bf B201}, 517 (1988). 
\bibitem{m89} E. Ma, Phys. Rev. Lett. {\bf 63}, 1042 (1989). 
\bibitem{ms18} E. Ma and U. Sarkar, Phys. Lett. {\bf B776}, 54 (2018).
\end{thebibliography}

\end{document}